# THE PHYSICS OF CORE-COLLAPSE SUPERNOVAE

*Supernovae are nature's grandest explosions and an astrophysical laboratory in which unique conditions exist that are not achievable on Earth. They are also the furnaces in which most of the elements heavier than carbon have been forged. Scientists have argued for decades about the physical mechanism responsible for these explosions. It is clear that the ultimate energy source is gravity, but the relative roles of neutrinos, fluid instabilities, rotation and magnetic fields continue to be debated.*

## Stan Woosley and Thomas Janka

Stan Woosley is in the Department of Astronomy and Astrophysics, University of California at Santa Cruz, Santa Cruz, California 95064, USA.
e-mail: woosley@ucolick.org

Thomas Janka is at the Max Planck Institute for Astrophysics,
Karl-Schwarzschild-Str. 1, D-85741 Garching, Germany.
e-mail: thj@mpa.garching.de

Few events in nature match the grandeur of supernovae. None surpasses their raw power: about $10^{53}$ erg s$^{-1}$ (equivalent to $10^{46}$ J s$^{-1}$) is released as neutrinos from a 'core-collapse' supernova, which is as much instantaneous power as all the rest of the luminous, visible Universe combined. Viewed on a cosmic scale, supernovae light up galaxies with spectacular fireworks that stir the interstellar and intergalactic media. They make most of the elements of nature, including those that form our own planet and bodies, and they give birth to the most exotic states of matter known — neutron stars and black holes. Supernovae have been at the forefront of astronomical research for the better part of a century, and yet no one is sure how they work.

From the outset, one must distinguish two kinds of supernovae, corresponding to two kinds of star death: Type Ia, thought to be the thermonuclear explosions of accreting white dwarf stars (1); and all the rest (Type II, Ib, Ic, and so on), which happen when the iron core of a massive star collapses to a neutron star or black hole. Observationally, Type I is defined by a lack of hydrogen lines in its spectrum, lines that Type II has. Type Ia supernovae happen in all types of galaxies with no preference for star-forming regions, consistent with their origin from an old or intermediate age stellar population. The rest happen only in star-forming regions where young massive stars are found. Here we will discuss just the latter variety, so-called core-collapse supernovae — the most frequent kind of supernovae in nature.

CORE COLLAPSE: THE GRAVITY BOMB

Since 1939, when Baade and Zwicky first suggested that supernovae are energized by the collapse of an ordinary star to a neutron star (2), scientists have tried to understand in detail how they work. The starting point is a star heavier than about 8 solar masses that has passed through successive stages of hydrogen, helium, carbon, neon, oxygen and silicon fusion in its centre (Table 1). With the passing of each stage, the centre of the star grows hotter and more dense. Indeed, the evolution of the inner parts of a massive star can be thought of as just one long contraction, beginning with the star's birth, burning hydrogen on the main sequence, and ending with the formation of a neutron star or black hole (Fig. 1). Along the way, the contraction 'pauses', sometimes for millions of years, as nuclear fusion provides the energy necessary to replenish what the star is losing to radiation and neutrinos. Each time one fuel runs out, the star contracts, heats up and then burns the next one, usually the ashes of the previous stage. After helium burning, the evolution is greatly accelerated by neutrino losses. For temperatures approaching a billion degrees or more, a large thermal population of electrons and positrons is maintained. When the electrons meet and annihilate with positrons, a neutrino–antineutrino pair is occasionally produced. These neutrinos escape the star with ease and force the burning to go faster to replenish the loss. Although the fusion of hydrogen and helium takes millions of years, the last burning phase — silicon burning — lasts only two weeks.

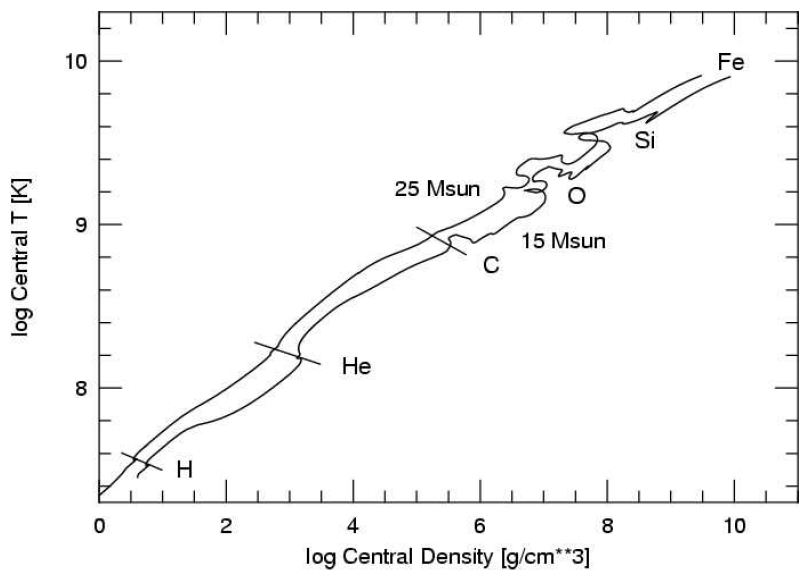

**Figure 1**: *The evolution of the temperature and density for the centre of two massive stars, 15 and 25 times heavier than the Sun. Labels show the location where the star pauses to burn a given fuel (Table 1). Overall, the evolution of a massive star is a continued contraction to higher density and temperature, a contraction that only ends when a neutron star or black hole is formed. During most of the evolution the density is proportional to the cube of the temperature, as expected for an ideal gas in hydrostatic equilibrium, but there are deviations caused by nuclear burning and the partial quantum mechanical degeneracy of the electrons.*

Eventually, a core of about 1.5 solar masses of iron-group elements is produced. Because the nuclear binding energy per nucleon has its maximum value for the iron group, no further energy can be released by nuclear fusion, yet the neutrino losses continue unabated, exceeding the Sun's luminosity by a factor of about $10^{15}$. At such high temperatures and densities, two other processes also rob the iron core of the energy it needs to maintain its pressure and avoid collapse — electron capture by nuclei, and an endoergic process called photodisintegration. At densities above $10^{10}$ g cm$^{-3}$, electrons are squeezed into iron-group nuclei, raising their neutron number. As electrons supply most of the pressure that holds the star up, their loss robs the core of both energy and support. At the same high temperature, radiation also begins to melt down some of the iron nuclei to helium — this is photodisintegration — partially undoing the last million years or so of nuclear evolution and sapping the core of still more energy. Soon the iron core is falling nearly freely at about a quarter of the speed of light. Starting from the size of the Earth, the core collapses to a hot, dense, neutron-rich sphere about 30 km in radius, a proto-neutron star. Eventually the repulsive component of the short-range nuclear force halts the collapse of the inner core when the density is nearly twice that of the atomic nucleus, or $4\text{–}5 \times 10^{14}$ g cm$^{-3}$.

TABLE 1 Evolution of a 15-solar-mass star.

| Stage | Time Scale | Fuel or Product | Ash or product | Temperature ($10^9$ K) | Density (gm/cm$^3$) | Luminosity (solar units) | Neutrino Losses (solar units) |
|---|---|---|---|---|---|---|---|
| Hydrogen | 11 My | H | He | 0.035 | 5.8 | 28,000 | 1800 |
| Helium | 2.0 My | He | C,O | 0.18 | 1390 | 44,000 | 1900 |
| Carbon | 2000 y | C | Ne,Mg | 0.81 | $2.8 \times 10^5$ | 72,000 | $3.7 \times 10^5$ |
| Neon | 0.7 y | Ne | O,Mg | 1.6 | $1.2 \times 10^7$ | 75,000 | $1.4 \times 10^8$ |
| Oxygen | 2.6 y | O,Mg | Si,S,Ar,Ca | 1.9 | $8.8 \times 10^6$ | 75,000 | $9.1 \times 10^8$ |
| Silicon | 18 d | Si,S,Ar,Ca | Fe,Ni,Cr,Ti,... | 3.3 | $4.8 \times 10^7$ | 75,000 | $1.3 \times 10^{11}$ |
| Iron core collapse[a] | ~1 s | Fe,Ni,Cr, Ti,... | Neutron Star | > 7.1 | $>7.3 \times 10^9$ | 75,000 | $>3.6 \times 10^{15}$ |

[a]The presupernova star is defined by the time when the contraction speed anywhere in the iron core reaches 1,000 km s$^{-1}$.

The abrupt halt of the collapse of the inner core and its rebound generates a shock wave as the core's outer half continues to crash down. Once it was thought that this bounce might actually be the origin of the supernova's energy (3,4), that the outward velocity of the bounce would grow as it moved into the outer layers of the core and eject the rest of the star with high velocity. Now it is known that this does not occur. Instead, the shock wave stalls due to photodisintegration and copious neutrino losses. A few milliseconds after the bounce, all positive velocities are gone from the star and the dense, hot neutron-rich core (commonly called a proto-neutron star) is accreting mass at a few tenths of a solar mass per second. If this accretion continued unabated for even one second, the proto-neutron star would be crushed into a black hole and no supernova would ever explode.

However, the proto-neutron star emits a prodigious luminosity of neutrinos. Over the next few seconds, if it does not become a black hole, it will radiate about 10% of its rest mass (about $3 \times 10^{53}$ erg), eventually settling down as a gigantic neutron-rich nucleus of 10-km radius — a neutron star. This neutrino emission is actually the chief output of the event which is overwhelmingly a gravity-powered neutrino explosion. But how can this be used to turn the collapse of the rest of the star into the explosion that we see with optical telescopes? This is the part of the problem that has caused theorists the greatest difficulty for forty years (5). A typical core-collapse supernova has $1\text{–}2 \times 10^{51}$ erg in kinetic energy, far less than that released as neutrinos during neutron-star formation. But the neutrinos streaming out from the core have a small cross-section for energy deposition and, to make matters worse, a large part of the energy they do deposit is radiated away again as neutrinos (neutrinos deposit their energy chiefly by the reactions $p + \bar{\nu} \rightarrow n + e^+$ and $n + \nu \rightarrow p + e^-$, where $p$, $n$, $e^+$, and

$e^-$ are the proton, neutron, positron and electron respectively; they are radiated away by the inverse of these same reactions). The efficiency for absorption and re-emission depends upon the density and temperature structure around the neutron star, and this, in turn, depends upon some complicated fluid mechanics (6). There is also a threshold of energy that must be deposited in a brief time to overcome the 'ram pressure' of the infalling matter, which, as we noted, is rapidly accreting (7).

The current frontier in in research into core-collapse supernovae centres on multi-dimensional simulations of the contracting proto-neutron star and neutrino energy deposition in its immediate surroundings. If this neutrino-powered model is to work, neutrino energy deposition must inflate a large bubble of radiation and electron–positron pairs surrounding the neutron star. The outer boundary of this inflating bubble becomes the outgoing shock wave that ejects the rest of the star and makes the explosion (Fig. 2). Because the entropy is high at the base of this bubble, it is convectively unstable. This convection has salutary effects: it cools the regions where the neutrinos are depositing their energy and thus reduces subsequent losses; and it carries the energy deposited in a small region to large radii where it can work effectively against the infalling matter at the shock. Only recently has it been realized that convection also conspires with a generic instability of the accreting shock against non-radial deformation (8,9). This leads to asymmetries whose dominant mode, in non-rotating cases, can be dipolar(10, 11) (Fig. 3). Matter flows in on one side of the proto-neutron star, is heated by neutrino energy deposition, and flows out on the other.

This 'jet engine' aspect of the explosion can only be seen in calculations that carry at least half of the whole core on the grid (many previous calculations only examined quadrants), and it might offer a physical explanation for the so called 'kicks' observed in young neutron stars. Typical pulsars are observed to be travelling with a large random velocity (12), averaging 300–400 km s$^{-1}$ (values of more than 1,000 km s$^{-1}$ are sometimes seen). Even with no rotation, the dipole pattern of convection breaks the spherical symmetry and provides a preferred axis. This axis is chosen randomly — an example of spontaneous symmetry breaking. Convective motion and asymmetric mass ejection then lead to a kick of the neutron star in the opposite direction. The expanding debris of the exploding star is deformed and its composition stirred up, with elements made deep inside mixed to the outside — as seen, for example, in the Cassiopeia A supernova remnant (13). The calculations also give acceptable remnant neutron-star masses (about 1.4 solar masses) and explosion energies for a particular parameterization of the contraction and neutrino emission of the core of the proto-neutron star.

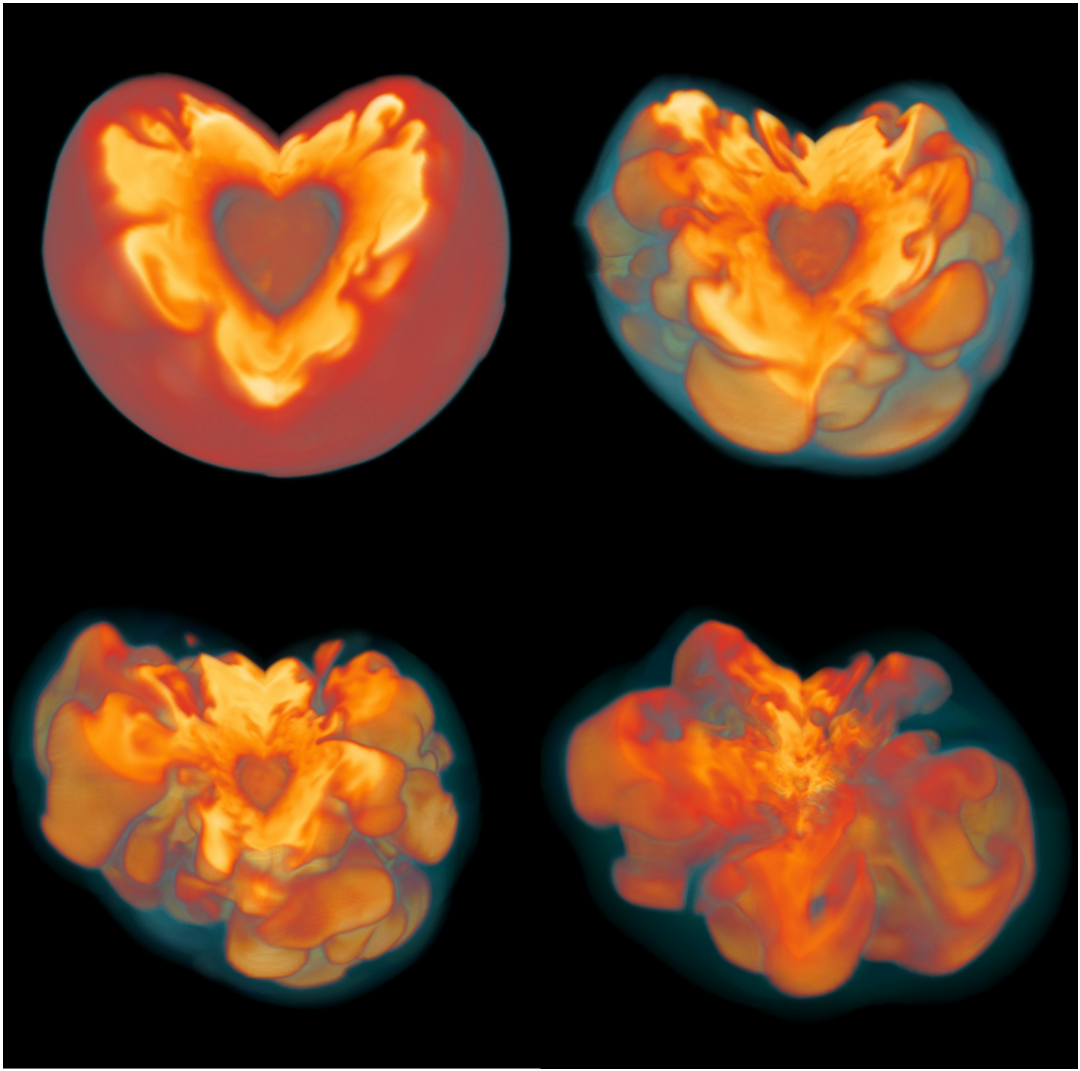

**Figure 2:** *Looking into the heart of a supernova (14). Four snapshots show the vigorous boiling of the neutrino-heated, convective region around the nascent neutron star. Buoyant bubbles of hot matter moving outwards appear bright red and yellow. These are bounded by a shock wave, which expands outwards, disrupting the star. The images, from top left to bottom right, show the structure at 0.1, 0.2, 0.3, and 0.5 seconds after the shock is born. At these times, the shock has an average radius of about 200, 300, 500, and 2,000 kilometers, respectively.*

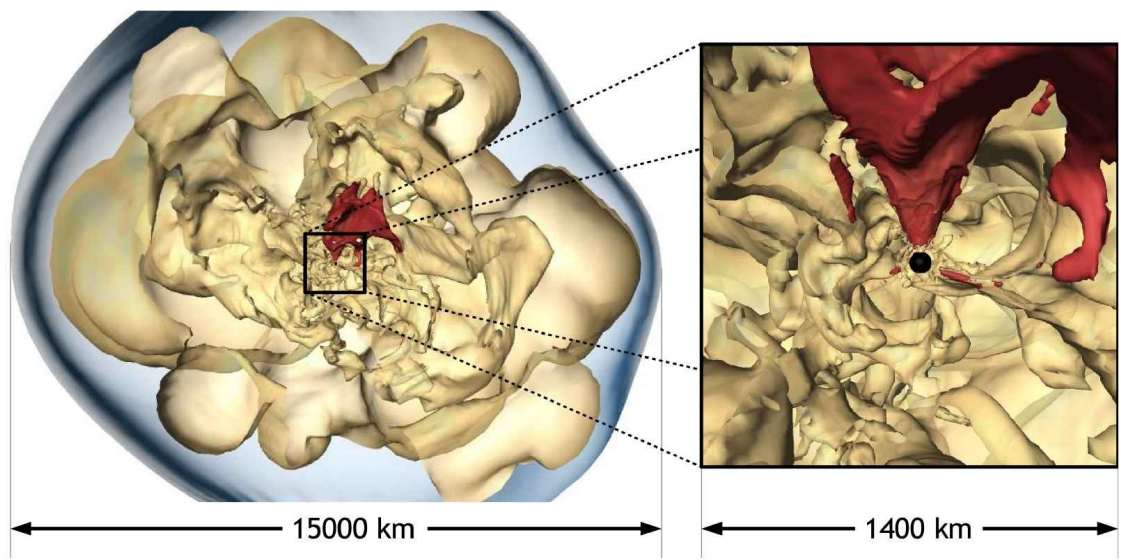

**Figure 3:** *Accretion onto the nascent neutron star shows a dipolar character (15). Cool matter (visible in red in the blow-up on the right) falls and is funnelled onto one side of the neutron star (black circle at the center), while neutrino-heated, hot ejecta flows out on the other. This 'jet engine' can accelerate the neutron star to velocities of several hundred kilometres per second within the first second of its life. At that same time, the supernova shock wave (blue, enveloping surface) is already well on its way through the exploding star (left panel), being pushed by the buoyant bubbles of neutrino-heated gas. Although the calculation was followed in three spatial dimensions, the initial model was spherically symmetric and was not rotating.*

However, parameter-free multi-dimensional models, with neutrino transport included consistently throughout the entire mass, yield ambiguous results on the key issue of whether the star actually explodes. At least four groups worldwide are currently attacking this problem with two- and three-dimensional simulations that tax the world's fastest supercomputers. The problem is very difficult. Not only must the whole neutron star be carried in a simulation with high resolution, but the transport and interactions of six varieties of neutrinos (electron-, muon- and tau-neutrinos and their anti-particles) must be followed. The resolution must be high, so as not to make the numerical viscosity too great to follow the convection properly. Three spatial dimensions is preferable to two because the convection may have different properties if cylindrical symmetry is enforced. The neutrino radiation, which is non-thermal, may have important angular structure as it moves from the 'optically thick' to transparent regions. Many energy groups and angle groups must be carried in calculations with of order one billion spatial mesh points. The different groups studying the problem use different approximations and, to this point, no one has carried out a definitive three-dimensional simulation using neutrino physics and resolution that give the world-wide community confidence in the results. Those calculations that have been carried out in various approximations give results ranging from vigorous explosions (16,17), to marginal failures (18), to outright failures (19).

But there is good reason to hope that this 65-year theoretical odyssey might be coming to a conclusion. With recent advances in computer technology and the worldwide development of the necessary multi-dimensional radiation-hydrodynamics codes, a definitive calculation of the simplest non-rotating, non-magnetic model using standard neutrino physics should be within our grasp in the next few years. Several groups are gearing up for the task (18,20,21,22). These results will either resolve a long standing problem about how massive stars die or show that new physics is needed. Both are exciting prospects. The new physics could include the effects of rotation and magnetic fields, revisions to the high-density equation of state used to describe the neutron star interior, or changes to neutrino physics.

It should be noted however, that there is really not just one supernova problem but many, because the outcome depends on the mass of the star. More massive stars develop bigger iron cores, eventually so large that they hover on the cusp of becoming black holes when they collapse. The matter that surrounds the iron core in shells of silicon and oxygen has higher density in the more massive stars and accretes more rapidly. A successful outgoing shock is harder to get started. It may well be that the range of masses that actually explodes by forming a neutron star is limited (23). The probable existence of stellar mass black holes in many binaries (24) suggests that sometimes supernovae make black holes, not neutron stars. Conversely, stars on the lighter end of the mass range — stars of 8–11 solar masses — have outer layers that are very loosely bound and may be relatively easy to explode (25,26). These stars, however, do not eject enough mass to explain the origin of abundant heavy elements such as oxygen, magnesium, silicon, sulphur and calcium.

One of the major handicaps of supernova theory is the scarcity of direct observations of the events that transpire in the supernova core. One can learn indirectly from the light curves, spectra and the ejected chemical elements, but direct obervations are only possible using neutrinos or gravitational radiation. So far, observations of neutrinos from one supernova, SN 1987A, have shown that a neutron star formed of the anticipated binding energy, diffusion time scale, and approximate mass (27), but the number of detected neutrinos was small. Measurements from a future supernova in the Milky Way (28, 29) will provide more information, including limits on the rotation rate. Large underground experiments to capture these ghostly and ethereal signals are under construction at various places around the globe.

ROTATION AND GAMMA-RAY BURSTS

The 'standard model' for core-collapse supernovae that has been described either ignores rotation or considers it an afterthought in an explosion principally energized by neutrinos. This assumption has never been universally accepted (30–33) and recent observations of gamma-ray bursts (GRBs) in simultaneous conjunction with supernovae (34,35) have shown that, at least in some cases, it is demonstrably false.

GRBs — intense flashes of gamma-rays lasting about 20 seconds and coming from cosmological distances — are produced by highly relativistic collimated outflows (with Lorentz factors, $\Gamma = (1-(u/c)^2)^{-1/2} > 200$, where $u$ is the outflow velocity) (36). A typical GRB involves a jet with an opening angle of about 5 degrees (37). Supernovae have been seen in conjunction with GRBs on at least three occasions and it may be that they accompany all GRBs of the so-called 'long soft' variety (the most common kind) (38). Those GRB-supernovae that have been studied in detail have unusual spectra and light curves suggesting a high mass and very high energy — about $10^{52}$ erg, or 10 times that of a typical supernova. The term 'hypernova' is sometimes applied. No spherically symmetric model, even with an energy of $10^{52}$ erg, is capable of accelerating sufficient mass to the necessary high speeds, and any spherical explosion is inconsistent with the beaming observed in GRBs. These events, which must also be the deaths of massive stars, have a broken symmetry.

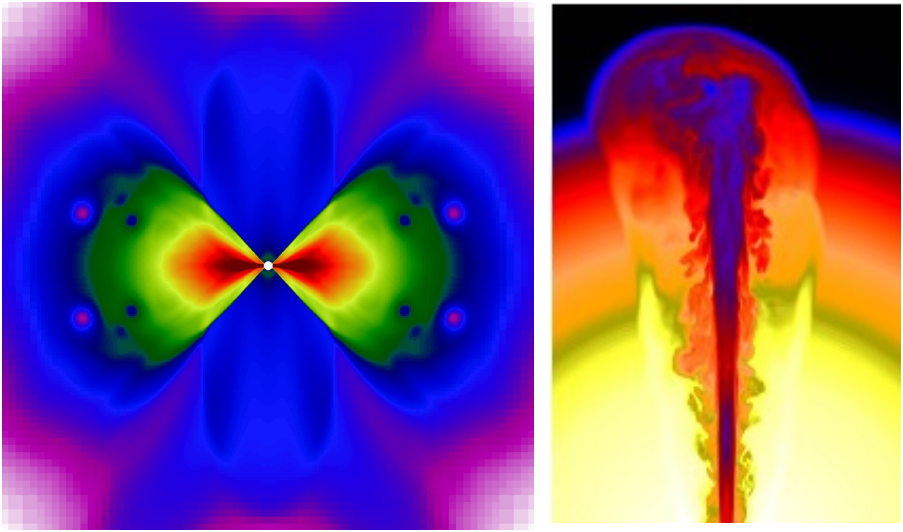

**Figure 4:** *The collapse of a the core of a rapidly rotating 14 solar mass helium core yields a black hole and a centrifugally supported accretion disk (42). **a,** This image, representing an area 1,800 km across, shows the density structure 20 seconds after the black hole has formed and begun to accrete. At this point, the black hole mass is 4.4 solar masses, corresponding to a radius of 14 km, and the accretion rate has been 0.1 solar masses per second for the last 15 seconds. The highest densities (dark red) are about $10^9$ g cm$^{-3}$; the lowest along the axis near the hole, $10^7$ g cm$^{-3}$. In this collapsar model, jets initiated either by magnetic processes in the disk or by the rotating black hole itself propagate up the axis, exit the star and eventually produce gamma-ray bursts. **b,** A jet exiting the star, which has a radius similar to that of the Sun. The time here is 8 seconds after the jet originated in the centre.*

All credible models of long-soft GRBs so far rely on very rapid rotation to produce either a neutron star rotating nearly at the point of centrifugal break up (39,40), or a black hole and an accretion disk (41) — a 'collapsar'. The rotational axis gives a natural preferred direction for the propagation of a jet (42) (Fig. 4). Neutrinos play a role in some of these models, but the observed supernova energy is much greater than what they are able to transport. It seems certain that long-soft GRBs are rotationally powered supernovae. However, even correcting for the fact that we only see a GRB if we are in the solid angle of its beam, GRBs are associated with but a small fraction of all supernovae, roughly a few tenths of a percent (43). The physics that drives them therefore need not necessarily be the same as for ordinary supernovae. But if it is not, what causes a star to follow one path towards death rather than another? Is there a continuum of events, some in which rotation dominates, some in which it is negligible, and others in between? Or is rotation (and the accompanying magnetic fields that differential rotation generates) a necessary component of all supernova explosions?

The answer may lie in the evolution of a star when it was younger (44) and perhaps even the characteristics of its formation — especially its composition and angular momentum. Massive stars with a composition like that of the sun spend a phase of their lives as red supergiants. During that phase, their outer layers expand by a factor of several hundred and, no matter what their initial angular momentum, rotate very slowly. The shear between this slowly rotating envelope and the rapidly rotating, denser core tends to slow down the rotation of the latter. Magnetic fields that thread both the envelope and core may be particularly efficient at doing this (45). If the star does not lose its envelope rapidly, either to a wind or a binary companion, it will be slowed too much to emit a GRB. Instead, it will have an explosion where rotation plays, at most, a secondary role and it will leave behind a pulsar rotating at a typical rate, with a period of 10–20 ms. This is probably the path taken by most massive stars (46).

If the star is able to bypass red-giant formation — as it might if it rotates fast enough on the main sequence (47,48), or if it quickly loses its envelope to a companion star — the core will retain its rapid rotation. But this is only the first hurdle, because stripped-down helium stars, called Wolf-Rayet stars, are known to lose mass at a rapid rate. Mass loss carries away the angular momentum and, for currently accepted loss rates, Wolf-Rayet stars of solar composition would not retain enough angular momentum to become a GRB. However, the mass-loss rate depends on the metal content of the primordial star, especially the fraction of iron (49). The many atomic lines of iron trap momentum from the outgoing radiation causing the surface layers to blow away. Low-metallicity Wolf-Rayet stars, deficient in iron, may retain enough mass and angular momentum to make a GRB provided their metallicity is less than about 20% that of the Sun (47).

Much of this picture is controversial and still to be fleshed out, but it does point to a tantalizing synthesis in which the supernovae that are most common today, and the neutron stars they leave behind, are the result of neutrino-powered explosions. The resulting pulsars rotate slowly and have large velocities. But as the heavy-element content of the gas-making stars goes down, an increasing fraction of the most rapidly rotating stars would die a different sort of death, one in which magnetic fields played a major role. Black hole formation would be common and collapsars and rapidly rotating neutron stars would be more abundant — although still a small fraction of all massive-star deaths. In the distant past, before supernovae had contaminated the pure hydrogen and helium of the Big Bang with heavier elements, the metallicity was quite low. Although islands of low metallity remain today — for example, in galaxies such as the Small Magellanic Cloud — violent events like gamma-ray bursts would be chiefly a phenomenon of the young universe.

ELEMENT PRODUCTION BY THE R-PROCESS

Roughly one half of the isotopes heavier than the iron group show evidence of having been produced on a very rapid time scale (of the order of one second) in a high temperature environment by a very high flux of neutrons (more than $10^{20}$ neutrons per cm$^3$, at temperatures greater than $10^9$ K)(50). These are referred to as the r-process nuclei (where 'r' stands for rapid neutron capture). The necessary conditions can only be achieved in explosive situations, and it has long been thought that supernovae are the most likely production site, although a frequently mentioned alternative is merging neutron stars (51). The latter is probably ruled out as the dominant r-process source, since the low rates of occurrence would lead to r-process enrichment that is not consistent with observations (52). Neutron-rich winds or jets from GRB supernovae, which are discussed as yet another potential r-process source (for example, in ref. 53), might face similar problems.

Within supernovae, the necessary conditions to achieve solar abundances through the r-process exist only in the innermost ejecta, closest to the neutron star (Fig. 5). There, the intense flux of neutrinos flowing out of the cooling neutron star blows a neutron-rich 'wind', which lasts about 10 seconds. The explosion has already ejected most of the stellar material in the near vicinity and it is thought that the composition of the wind reflects chiefly the properties of the compact object rather than those of the star that made it. The production of the r-process is therefore primary, in the sense that the same set of relative abundances would result from a neutron star produced long ago by a star with a low initial concentration of heavy elements as from a star today. Observations are consistent with this result. Very old, metal deficient stars display the same elemental ratios for their heavy elements as the r-process abundance ratios in the Sun (54). Because r-process nucleosynthesis depends upon the properties of the neutron star when the supernova was only a few seconds old, it is a powerful constraint on the physics of the explosion. It directly samples the neutrino luminosities and temperatures.

To see this, one must delve more deeply into how the r-process works. As electron neutrinos and their anti-particles stream out, some of them interact with matter in the neutron star's atmosphere. At such early times, the atmosphere is so hot that it consists of radiation, electron–positron pairs and unbound neutrons and protons — much like the early Universe. Each interaction of a neutrino and a neutron makes a proton and an electron. Conversely, antineutrinos plus protons make neutrons and positrons. A given nucleon will interact many times with both neutrinos and antineutrinos before gaining sufficient energy to escape, so the steady-state ratio of neutrons to protons comes to reflect the ratio of the fluxes of the two kinds of neutrinos and their energies (the cross-section for interaction with neutrinos goes as the neutrino energy squared). After one second, the fluxes of neutrinos and antineutrinos are almost the same, but the antineutrinos are hotter because the outer layers of the neutron star are neutron-rich and antineutrinos pass through these layers with greater ease than neutrinos do. Because of this higher mean energy of the antineutrinos, the steady-state abundances in the wind favour an excess of neutrons compared with protons.

As the wind flows out, it cools. Starting at about 10 billion K, the nucleons begin to combine with one another to make $^4$He nuclei, also known as α-particles. Each α-particle formed removes two neutrons and two protons from the wind, but since there are fewer protons than neutrons, this eventually results in a wind consisting mostly of α-particles and free neutrons. Neutrons do not capture on α-particles since $^5$He is very unstable. But, at 5 billion K, some of the α-particles start to assemble into heavier nuclei, especially those in the iron group. This leaves a mixture of (mostly) α-particles, neutrons and a few iron-group nuclei. During the next seconds, as the temperature cools to 1 billion K, hundreds of free neutrons are captured by these few heavy nuclei, producing the *r*-process (Fig. 5).

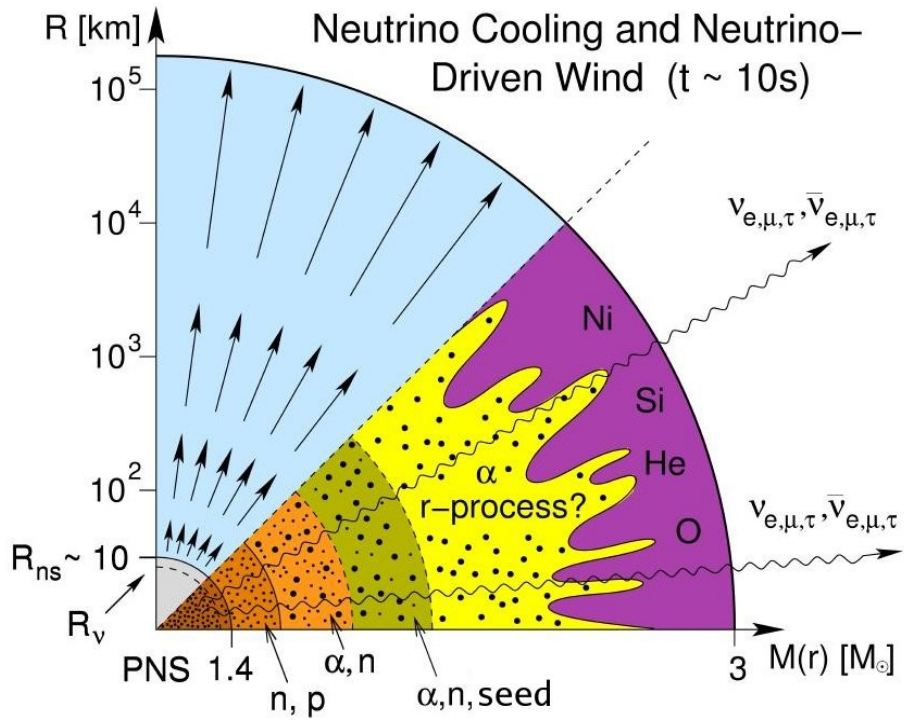

**Figure 5:** *Neutrinos ($\nu_e, \nu_\mu, \nu_\tau$ and their anti-particles) drive a wind from the surface of the cooling proto-neutron star (PNS) creating the r-process isotopes. The wind begins as a flux of neutrons and protons lifted from the surface of the PNS (here 1.4 solar masses and 10 km in radius) by neutrinos originating at the "neutrinosphere" ($R_\nu$). As these nucleons flow out, an excess of neutrons is created by the capture of anti-neutrinos on protons. As the nucleons cool, all the available protons combine with neutrons to make $\alpha$-particles until one is left, in the orange region, with a mixture of only $\alpha$-particles and unbound neutrons. Further cooling leads to the assembly of a few $\alpha$-particles into nuclei in the iron group ("seed") by reactions involving neutrons and $\alpha$-particles (green region). As the temperature declines still further, from 3 billion K to 1 billion K, all neutrons are captured on this seed making the heavy r-process nuclei. Since the efficiency of the reactions that assemble $\alpha$-particles into seed increases with the density, lower density in the wind keeps the number of seed small and makes the number of neutrons that can be captured on each larger.*

When it works, it works well (55). About $10^{-5}$ solar masses of wind material is ejected and 10 – 20% of that is r-process (the rest is helium). This is what is needed when the ejecta of the $10^8$ supernovae that have happened in our Galaxy are all added up, mixed with the rest of the gas, and compared with abundances found in the Sun (which collapsed out of this gas). Unfortunately, the devil is in the details and when carefully calculated, the wind properties do not quite give what is needed. At a given temperature, the wind is about 4 times too dense. With the higher density, too many $\alpha$-particles reassemble into heavy nuclei, making the neutron-to-heavy-nuclei ratio far too small to make r-process nuclei with atomic mass 200.

Since the essential aspects of this model were laid out a decade ago, numerous studies have confirmed this basic result. In the simplest one-dimensional models with no rotation or magnetic fields, elements above atomic mass 100 are not made by the r-process. The relevant cross-sections are well determined, so something must be missing in the explosion model. One must either forsake this very promising site for the r-process or add something new.

The missing ingredients may, once again, be rotation and magnetic fields. Confining the outward flow of the wind with a magnetic field allows more neutrino capture, more energy deposition, and hence a lower density in the expanding material (56). Adding just a small amount of energy further out in the wind, after the mass-loss rate has already been determined, could also significantly reduce the density where the r-process occurs. Shortening the time scale by increasing the wind velocity also helps. Perhaps energy is put in by magnetic reconnection following the wrapping up of a seed field by rotation (57), or by acoustic waves(57,58) or Alfven waves from the neutron star (59). It is not clear if the necessary magnetic fields and rotation are so large as to have a direct influence on the explosion mechanism or to break its spherical symmetry, but something beyond the standard, non-rotating, non-magnetic model is strongly suggested.

CONCLUSION

Despite continued uncertainty regarding the fundamental question of whether non-rotating massive stars are able to explode by neutrino energy transport alone, progress is being made. In a few years, we should know whether standard physics can give explosions and explain the high peculiar velocities seen for pulsars without invoking magnetic fields, rotation or modified descriptions of basic neutrino physics. Even if they turn out to be unimportant to the explosion mechanism, rotation and magnetic fields may be essential to explain the r-process and are definitely crucial to understanding GRBs. The tantalizing possibility exists of a grand synthesis in which the most common supernovae are the aftermath of neutrino explosions in slowly rotating stars, whereas rare, hyper-energetic phenomena such as GRBs are phenomena at the other end of a spectrum of stellar evolution in which rotation is the key variable. We should know soon.

ACKOWLEDGEMENTS

This work was supported by the SciDAC Program of the US Department of Energy (DC-FC02-01ER41176), the National Science Foundation (AST 02-06111), NASA (NAG5-12036), and the German Research Foundation within the Collaborative Research Center for Astroparticle Physics (SFB 375) and the Transregional Collaborative Research Center for Gravitational Wave Astronomy (SFB-Transregio 7).